\documentclass[prd,nofootinbib,twocolumn]{revtex4}
\usepackage{amsmath}

\begin{document}

\title{The Kepler equation for inspiralling compact binaries}
\author{Zolt\'{a}n Keresztes, Bal\'{a}zs Mik\'{o}czi, L\'{a}szl\'{o} \'{A}.
Gergely}
\affiliation{Departments of Theoretical and Experimental Physics, University of Szeged,
Szeged 6720, Hungary}

\begin{abstract}
Compact binaries consisting of neutron stars / black holes on eccentric
orbit undergo a perturbed Keplerian motion. The perturbations are either of 
\textit{relativistic} origin or are related to the\textit{\ spin, mass
quadrupole and magnetic dipole moments }of the binary components\textit{.}
The post-Newtonian motion of such systems decouples into radial and angular
parts. We present here for the first time the radial motion of such a binary
encoded in a \textit{generalized Kepler equation,} with the inclusion of all
above-mentioned contributions, up to linear order in the perturbations.
Together with suitably introduced parametrizations, the radial motion is
solved completely.
\end{abstract}

\date{\today }
\maketitle
\startpage{1}

\section{Introduction}

The worldwide effort to capture gravitational wave signals emitted by
astrophysical sources is under way. A network of interferometric
gravitational wave detectors \cite{LIGO}-\cite{Tama} is either already
operational, or close to completion. Compact binaries consisting on neutron
stars / black holes are among the best candidates to emit gravitational
radiation in the bandwidth of these detectors. Upper limits on the
gravitational radiation emitted by such binaries were already found \cite%
{LIGO1}, \cite{LIGO2} from the S2 scientific run of LIGO.

The evolution of such a binary system can be divided into three phases:
inspiral, merger and ringdown. The merger phase can be understood only by
numeric simulations. Even in the last part of the inspiral a numerical
treatment seems adequate due to the intermediate binary black hole (IBBH)
problem \cite{IBBH}, studied also in \cite{Buonanno1}. Earlier, in the
inspiral phase a post-Newtonian (PN) description of high accuracy provides
satisfactory results. The PN corrections of relativistic nature are known to
3PN orders \cite{3PN}. However there are other contributions to be taken
into account, related to various physical characteristics of the binary
components as well.

For compact binaries, there is the spin-orbit (SO) interaction appearing at
1.5 PN orders. At this accuracy both the spin vectors $\mathbf{S}_{\mathbf{i}%
}$ and the orbital angular momentum $\mathbf{L}$ undergo a precessional
motion about the total angular momentum $\mathbf{J}$ \cite{BOC}. This is a
novel feature in the post-Newtonian evolution of the system. Such an effect
was recently claimed \cite{OC} to be observable for the J0737-3039A/B double
pulsar \cite{Burgay}, \cite{Lyne}.

The precessional motion of the spin(s) is called \textit{simple precession},
whenever the two masses are equal, or one of the spins can be neglected, say 
$\mathbf{S}_{\mathbf{2}}=0$ \cite{ACST}. These two cases were studied in 
\cite{KG}, where among other results, the Kepler equation was derived up to
3PN orders with the inclusion of the SO contributions. As a related result,
the evolution of the relativistic periastron advance parameter was recently
computed \cite{KG2}. The tilt angle of the spin with respect to $\mathbf{L}$
was estimated to be smaller than $\simeq 60^{\circ }$ from generic
astrophysical considerations on the evolution of compact binaries \cite{GIKB}%
.

The SO interaction gives corrections to the losses of energy and magnitude
of angular momentum of the system occurring due to gravitational radiation.
For eccentric orbits these were given by \cite{RS} and \cite{GPV}. An other
work relying on the use of the Effective One-Body approach \cite{BD99}, \cite%
{BD00}, has employed the SO contribution in the study of the inspiral to
plunge phase of the coalescence \cite{BCD05}.

Moreover, in the two cases of simple precession detection template families
have been worked out both containing a set of phenomenological parameters 
\cite{BCV2} or physical parameters \cite{BCV2P}. The latter would allow for
determining the angle $\kappa _{1}$ and the magnitude of the single spin (in
fact of $\chi =S_{1}/m_{1}^{2}$) from the study of gravitational radiation.

At 2PN an other set of new effects related to various physical
characteristics of the compact binary emerge. The losses of energy and
magnitude of angular momentum of the system on eccentric orbit, due to
gravitational radiation were derived in \cite{spinspin} for the spin-spin
interaction, in \cite{quadrup} for the mass quadrupole-monopole interaction
and in \cite{mdipol} for the interaction of magnetic dipoles.

Thus at 2PN physical quantities like the mass quadrupole and magnetic dipole
moments, as well as angular variables characterizing the spins and moments
appear in the formalism. Neither detection templates, nor methods to find
out these new physical parameters have been worked out so far. We note that
in principle, the observation of the evolution of gravitational wave
frequency \cite{MVG} allows to impose constraints on a \textit{combination}
of these parameters, but does not allow to predict their individual values.
There is still much to do until a complete understanding of the complicated
motion the system, occurring when all these interactions are taken into
account, will be achieved. Our present work fills an important gap in the
description of compact binaries with the enlisted physical characteristics.

In Section 2 we describe the post-Newtonian motion of such a binary system.
The radial part of the motion decouples and defines a radial orbit. We give
the generic expressions of the turning points of the radial motion which
allows for the introduction of the semimajor axis $a_{r}$ and radial
eccentricity $e_{r}$. We define the generalized true anomaly $\chi $ and
eccentric anomaly $\xi $ parameters. Our generalized true anomaly parameter
is different from the one employed in the Damour-Deruelle formalism and we
establish their relation. In Section 3 we derive the main result of the
paper, which is the generalized Kepler equation

\begin{eqnarray}
n\left( t-t_{0}\right) &=&\xi -e_{t}\sin \xi  \notag \\
&&+f_{t}\sin \left[ \chi +2\left( \psi _{0}-\overline{\psi }\right) \right] 
\notag \\
&&+\sum_{i=1}^{2}f_{t}^{i}\sin \left[ \chi +2\left( \psi _{0}-\psi
_{i}\right) \right] \text{ \ }.  \label{Kepler}
\end{eqnarray}%
The explicit expressions of the orbital elements $n$, $e_{t}$, $f_{t}$ and $%
f_{t}^{i}$ are given in Section 4, together with all contributions to $a_{r}$
and $e_{r.}$ They contain the relativistic contributions (PN-terms),
together with the SO, SS, DD and QM terms, all to linear order. One of the
consequences of taking into account the physical characteristics of the
binary, like the spins, mass quadrupole and magnetic dipole moments is the
emergence of novel angle variables in the Kepler equation (\ref{Kepler}).
The angles $\psi _{i}$ are the azimuthal angles of the spins and $2\overline{%
\psi }=\psi _{1}+\psi _{2}$. (For more details on the notations see \cite%
{GPV}.) The angle $\psi _{0}$ is the argument of the periastron (defined
here as the angle subtended by the periastron and the intersection line of
the planes perpendicular to the total and orbital angular momenta,
respectively).

We emphasize that in the description of the SO\ interaction several
spin-supplementary conditions (SSC) can be used. In order to simplify the
formalism, in this paper we use the non-covariant SSC of Pryce \cite{SSC1}
and Newton and Wigner \cite{SSC2}. In this SSC the Lagrangian is not
acceleration-dependent and the radial equation is simpler than in the
covariant SSC \cite{SSC1} employed earlier in \cite{GPV}.

\section{Generalized true and eccentric anomaly parametrizations}

The linear contributions to the motion of the compact binary can be
collected in the Lagrangian%
\begin{equation}
\mathcal{L}=\mathcal{L}_{N}+\mathcal{L}_{PN}+\mathcal{L}_{SO}+\mathcal{L}%
_{SS}+\mathcal{L}_{QM}+\mathcal{L}_{DD}\ ,
\end{equation}%
with the various contributions derived first in \cite{DD} (PN), \cite{KWW}
(SS), \cite{Poisson} (QM) and \cite{IT} (DD): 
\begin{eqnarray}
\mathcal{L}_{N} &=&\frac{\mu \mathbf{v}^{2}}{2}+\frac{Gm\mu }{r}\text{ }, 
\notag \\
\mathcal{L}_{PN} &=&\frac{1}{8c^{2}}\left( 1-3\eta \right) \mu v^{4}+\!\frac{%
Gm\mu }{2rc^{2}}\!\left[ \!\left( 3\!+\!\eta \right) v^{2}\!+\!\eta \dot{r}%
^{2}\!-\!\frac{Gm}{r}\!\right] \text{ },  \notag \\
\mathcal{L}_{SO} &=&\frac{G\mu }{2c^{2}r^{3}}\mathbf{v}\cdot \lbrack \mathbf{%
r}\times (4\mathbf{S}+3{\mbox{\boldmath $\sigma$}})]\text{ },  \notag \\
\mathcal{L}_{SS} &=&\frac{G}{c^{2}r^{3}}\left[ \left( \mathbf{S}_{\mathbf{1}%
}\cdot \mathbf{S}_{\mathbf{2}}\right) -\frac{3}{r^{2}}\left( \mathbf{r\cdot S%
}_{\mathbf{1}}\right) \left( \mathbf{r\cdot S}_{\mathbf{2}}\right) \right] 
\text{ },  \notag \\
\mathcal{L}_{QM} &=&\frac{G\mu m^{3}}{2r^{5}}\sum_{i=1}^{2}p_{i}\left[
3\left( \mathbf{\hat{S}}_{\mathbf{i}}\cdot \mathbf{r}\right) ^{2}-r^{2}%
\right] \text{ },  \notag \\
\mathcal{L}_{DD} &=&\frac{1}{r^{3}}\left[ 3(\mathbf{n\cdot d_{1}})(\mathbf{%
n\cdot {d}_{2}})-\mathbf{d_{1}\cdot {d}_{2}}\right] \text{ }.  \label{Lag}
\end{eqnarray}%
Note that the SO-part of the Lagrangian above was not given before and it is
valid when the spin-supplementary condition (SSC) of Pryce \cite{SSC1} and
Newton and Wigner \cite{SSC2} is chosen. We have verified that the SO-part
of the acceleration\ derived from $\mathcal{L}_{SO}$ agrees with the
expression (A1b) of \cite{Kidder}. The magnitude and direction of the spins
are denoted as $S_{i}$ and $\mathbf{\hat{S}}_{\mathbf{i}}$. The angle
subtended by them is $\gamma =\cos ^{-1}(\mathbf{\hat{S}_{1}\cdot \hat{S}%
_{2})}$. Here $\mathbf{S}=\mathbf{S}_{\mathbf{1}}+\mathbf{S}_{\mathbf{2}}$
and ${\mbox{\boldmath $\sigma$}=}\left( m_{2}/m_{1}\right) \mathbf{S}_{%
\mathbf{1}}+\left( m_{1}/m_{2}\right) \mathbf{S}_{\mathbf{2}}$. The
magnitude and direction of the magnetic dipole moments $\mathbf{d}_{\mathbf{i%
}}$ are denoted as $d_{i}$ and $\mathbf{\hat{d}_{i}}$. They subtend the
angle $\lambda =\cos ^{-1}(\mathbf{\hat{d}_{1}\cdot \hat{d}_{2})}$ with each
other.\ In a coordinate systems $\mathcal{K}$ with the axes $(\mathbf{\hat{c}%
,\hat{L}\times \hat{c},\hat{L}})$, where $\mathbf{\hat{c}}$ is the unit
vector in the $\mathbf{J\times L}$ direction, the polar angles $\kappa _{i}$
and $\psi _{i}$ of the spins are defined as $\mathbf{\hat{S}_{i}=}(\sin
\kappa _{i}\cos \psi _{i},\sin \kappa _{i}\sin \psi _{i},\cos \kappa _{i})$
(see \cite{GPV}). In the coordinate system $\mathcal{K}^{i}$ with the axes $(%
\mathbf{\hat{b}_{i},\hat{S}_{i}\times \hat{b}_{i},\hat{S}_{i}})$, where $%
\mathbf{\hat{b}_{i}}$ are the unit vectors in the $\mathbf{S_{i}\times L}$
directions, respectively, the polar angles $\alpha _{i}$ and $\beta _{i}$ of
the the magnetic dipole moments $\mathbf{{d}_{i}\,}$are $\mathbf{{\hat{d}}%
_{i}=}(\sin \alpha _{i}\cos \beta _{i},\sin \alpha _{i}\sin \beta _{i},\cos
\alpha _{i})$ (see \cite{mdipol}). The quadrupolar parameters (see \cite%
{quadrup}) are defined as $p_{i}=Q_{i}/m_{i}m^{2}$, where $Q_{i}$ is the
quadrupole-moment scalar \cite{Poisson} of the $\ i^{th}$ axially symmetric
binary component with symmetry axis $\mathbf{\hat{S}}_{\mathbf{i}}$. The
reduced mass is $\mu =m_{1}m_{2}/m$ and $\eta =\mu /m$.

From (\ref{Lag}) a radial equation can be derived%
\begin{equation}
\dot{r}^{2}\!=\!\frac{2E}{\mu }\!+\!\frac{2Gm}{r}\!-\!\frac{\overline{L}^{2}%
}{\mu ^{2}r^{2}}\!+\!\!\sum_{i=0}^{3}\!\frac{\delta A_{i}}{r^{i}}\!-\!\frac{2%
\overline{L}\delta L}{\mu ^{2}r^{2}}\!-\!\frac{2\delta E}{\mu }\text{ }.
\label{rad2}
\end{equation}%
Here $\overline{L}=\left( 1/2\pi \right) \int_{0}^{2\pi }L(\chi )d\chi $ is
the angular average of the magnitude of orbital angular momentum $L(\chi )$, 
$\chi $ being the true anomaly parameter. The explicit values of $\overline{L%
}$ in the case of spin-spin, quadrupole-monopole and magnetic dipole-dipole
interactions were computed in \cite{spinspin}, \cite{quadrup} and \cite%
{mdipol}. $\overline{A}$ is the magnitude of the Laplace-Runge-Lenz vector
characterizing a Keplerian motion with $E$ and $\overline{L}$. The
coefficients $\delta A_{i}$ in Eq. (\ref{rad2}) are$\ $constant PN
perturbations given in the Table \ref{PNconst}, which can be read from \cite%
{DD}. 
\begin{table}[h]
\caption{Various post-Newtonian constants in $\protect\delta A_{i}$.}
\label{PNconst}%
\begin{tabular}{|l|l|}
\hline
$\delta A_{0}$ & $3(3\eta -1)\frac{E^{2}}{c^{2}\mu ^{2}}$ \\ \hline
$\delta A_{1}$ & $2(7\eta -6)\frac{EGm}{c^{2}\mu }$ \\ \hline
$\delta A_{2}$ & $-2(3\eta -1)\frac{E\overline{L}^{2}}{c^{2}\mu ^{3}}+(5\eta
-10)\frac{G^{2}m^{2}}{c^{2}}$ \\ \hline
$\delta A_{3}$ & $(-3\eta +8)\frac{G^{2}m^{2}\overline{L}^{2}}{c^{2}\mu ^{2}}
$ \\ \hline
\end{tabular}%
\end{table}
The SO, SS, QM and DD contributions to $\delta L$ and $\delta E$ are
enlisted in the Tables \ref{delL} and \ref{delE}. The shorthand notations $%
\alpha _{DD}$ and $\beta _{DD}\!(2\chi )$ are defined in Eq. (\ref{notations}%
). 
\begin{table}[h]
\caption{Various contributions to $\protect\delta L$.}
\label{delL}%
\begin{tabular}{|l|l|}
\hline
$SO$ & $\frac{G\mu \overline{L}}{2c^{2}r}\sum\limits_{i=1,j\neq i}^{2}\!%
\frac{4m_{i}+3m_{j}}{m_{i}}S_{i}\cos \kappa _{i}$ \\ \hline
$SS$ & $-\frac{G\mu ^{2}}{2c^{2}\overline{L}^{3}}S_{1}S_{2}\sin \kappa
_{1}\sin \kappa _{2}\bigl\{2\overline{A}\cos \left[ \chi +2\left( \psi _{0}-%
\overline{\psi }\right) \right] $ \\ 
& $+\left( 3Gm\mu +2\overline{A}\cos \chi \right) \cos 2\left( \chi +\psi
_{0}-\overline{\psi }\right) \bigr\}$ \\ \hline
$QM$ & $\frac{G\mu ^{3}m^{3}}{4\overline{L}^{3}}\sum_{i=1}^{2}p_{i}\sin
^{2}\kappa _{i}\bigl\{2\overline{A}\cos \left[ \chi +2\left( \psi _{0}-\psi
_{i}\right) \right] $ \\ 
& $+\left( 3Gm\mu +2\overline{A}\cos \chi \right) \cos 2\left( \chi +\psi
_{0}-\psi _{i}\right) \bigr\}$ \\ \hline
$DD$ & $\frac{\mu ^{2}d_{1}d_{2}}{2\overline{L}^{3}}\left[ \!(3Gm\mu \!+\!4%
\overline{A}\cos \chi )\beta _{DD}\!\left( 2\chi \right) \!-\!\overline{A}%
\sin \chi \frac{d\beta _{DD}\!(2\chi )}{d\chi }\right] $ \\ \hline
\end{tabular}%
\end{table}
\begin{table}[h]
\caption{Various contributions to $\protect\delta E$.}
\label{delE}%
\begin{tabular}{|l|l|}
\hline
$SO$ & no contribution \\ \hline
$SS$ & $-\frac{GS_{1}S_{2}}{2c^{2}r^{3}}\bigl\{3\cos \kappa _{1}\cos \kappa
_{2}-\cos \gamma $ \\ 
& $-3\sin \kappa _{1}\sin \kappa _{2}\cos 2\left( \chi +\psi _{0}-\overline{%
\psi }\right) \bigr\}$ \\ \hline
$QM$ & $\frac{G\mu m^{3}}{2r^{3}}\!\!\sum_{i=1}^{2}p_{i}\left[ 1-\!3\!\sin
^{2}\!\kappa _{i}\cos ^{2}\!\left( \!\chi \!+\delta _{i}\right) \right] $ \\ 
\hline
$DD$ & $\frac{d_{1}d_{2}}{2r^{3}}\left[ \alpha _{DD}-3\beta _{DD}\left(
2\chi \right) \right] $ \\ \hline
\end{tabular}%
\end{table}
\qquad

In \cite{param} a generic scheme was introduced for parametrizing such
perturbed Keplerian motions. The advantage of the generalized eccentric
anomaly parametrization and generalized true anomaly parametrization is that
a simple technique based on the residue theorem can be applied for computing
secular effects \cite{param}. The scheme was applied individually for each
of the SO, SS, QM, DD perturbations in \cite{GPV}, \cite{spinspin}, \cite%
{quadrup} and \cite{mdipol}, respectively. It is straightforward to derive
the PN contribution to these parametrizations. We give here in a concise
form both parametrizations, with the inclusion of the PN contribution as
well. Both are defined in terms of the turning points $r_{\min }$ and $%
r_{\max }$ of the radial motion, given by $\dot{r}=0$: 
\begin{eqnarray}
r_{{}_{{}_{\min }^{\max }}}\!\!\! &=&\!\!\frac{Gm\mu \!\pm \!\overline{A}}{%
-2E}\!+\!\delta r_{{}_{{}_{\min }^{\max }}}^{PN}\!+\!\delta r_{{}_{{}_{\min
}^{\max }}}^{SO}\!+\!\delta r_{{}_{{}_{\min }^{\max }}}^{SS}\!+\!\delta
r_{{}_{{}_{\min }^{\max }}}^{QM}\!+\!\delta r_{{}_{{}_{\min }^{\max }}}^{DD}%
\text{ },  \notag \\
\delta r_{{}_{{}_{\min }^{\max }}}^{PN}\!\! &=&\!\left( \eta -7\right) \frac{%
Gm}{4c^{2}}\pm \left( \eta +9\right) \frac{G^{2}m^{2}\mu }{8\overline{A}c^{2}%
}\mp \left( 3\eta -1\right) \frac{\overline{A}}{8\mu c^{2}}\text{ },  \notag
\\
\delta r_{{}_{{}_{\min }^{\max }}}^{SO}\!\! &=&-\!\frac{G\mu }{2c^{2}%
\overline{L}\overline{A}}(\overline{A}\!\mp \!Gm\mu
)\!\sum\limits_{i=1,j\neq i}^{2}\!\frac{4m_{i}+3m_{j}}{m_{i}}S_{i}\cos
\kappa _{i}\text{ },  \notag \\
\delta r_{{}_{{}_{\min }^{\max }}}^{SS}\!\! &=&\!-\frac{G\mu S_{1}S_{2}}{%
2c^{2}\overline{L}^{2}\overline{A}}\left[ (\overline{A}\mp Gm\mu )\alpha
_{SS}+\overline{A}\beta _{SS}\right] \text{ },  \notag \\
\delta r_{{}_{{}_{\min }^{\max }}}^{QM}\!\! &=&\!\frac{G\mu ^{2}m^{3}}{4%
\overline{L}^{2}\overline{A}}\sum\limits_{i=1}^{2}p_{i}\left[ (\overline{A}%
\mp Gm\mu )\alpha _{QM}^{i}+\overline{A}\beta _{QM}^{i}\right] \text{ }, 
\notag \\
\delta r_{{}_{{}_{\min }^{\max }}}^{DD}\!\! &=&\!\frac{\mu d_{1}d_{2}}{2%
\overline{L}^{2}\overline{A}}\left\{ (\overline{A}\mp Gm\mu )\alpha _{DD}+%
\overline{A}\beta _{DD}\right\} \text{ }.
\end{eqnarray}%
We have introduced the shorthand notations%
\begin{eqnarray}
\alpha _{SS} &=&3\cos \kappa _{1}\cos \kappa _{2}-\cos \gamma \text{ },\text{
}  \notag \\
\beta _{SS} &=&\sin \kappa _{1}\sin \kappa _{2}\cos 2\left( \psi _{0}-%
\overline{\psi }\right) \text{ },  \notag \\
\alpha _{QM}^{i} &=&2-3\sin ^{2}\kappa _{i}\text{ },\text{ }  \notag \\
\beta _{QM}^{i} &=&\sin ^{2}\kappa _{i}\cos 2\left( \psi _{0}-\psi
_{i}\right) \text{ },  \notag \\
\alpha _{DD} &=&2\cos \lambda +3(\rho _{1}\sigma _{2}-\rho _{2}\sigma
_{1})\sin \Delta \psi   \notag \\
&&-3(\rho _{1}\rho _{2}+\sigma _{1}\sigma _{2})\cos \Delta \psi \text{ }, 
\notag \\
\beta _{DD}(k\chi ) &=&(\sigma _{1}\sigma _{2}-\rho _{1}\rho _{2})\cos \left[
k\chi +2\left( \psi _{0}-\overline{\psi }\right) \right]   \notag \\
&&-(\rho _{1}\sigma _{2}+\rho _{2}\sigma _{1})\sin \left[ k\chi +2\left(
\psi _{0}-\overline{\psi }\right) \right] \text{ },  \notag \\
\beta _{DD} &=&\beta _{DD}(0)\text{ },  \label{notations}
\end{eqnarray}%
where%
\begin{eqnarray}
\rho _{i} &=&\sin \alpha _{i}\cos \beta _{i}\ ,  \notag \\
\sigma _{i} &=&\cos \alpha _{i}\sin \kappa _{i}+\sin \alpha _{i}\sin \beta
_{i}\cos \kappa _{i}\ .
\end{eqnarray}%
The generalized eccentric anomaly parametrization $r\left( \xi \right) $ is
then defined as%
\begin{equation}
r\left( \xi \right) =a_{r}\left( 1-e_{r}\cos \xi \right) \ .  \label{rxi}
\end{equation}%
The eccentric anomaly $\xi $ reduces to the eccentric anomaly parameter $u$
of \cite{DD} for 1PN perturbations. In Eq. (\ref{rxi}) the semimajor axis $%
a_{r}$ and the radial eccentricity$\ e_{r}$ was introduced as 
\begin{eqnarray}
a_{r} &=&\frac{r_{\max }+r_{\min }}{2}\text{ },  \label{ar} \\
e_{r} &=&\frac{r_{\max }-r_{\min }}{r_{\max }+r_{\min }}\text{ }.  \label{er}
\end{eqnarray}%
The generalized true anomaly parametrization is defined as:%
\begin{equation}
\frac{2}{r\left( \chi \right) }=\left( \frac{1}{r_{\min }}+\frac{1}{r_{\max }%
}\right) +\left( \frac{1}{r_{\min }}-\frac{1}{r_{\max }}\right) \cos \chi 
\text{ }.  \label{rchi}
\end{equation}%
These two parametrizations of the radial motion are interrelated by the
Keplerian relations%
\begin{eqnarray}
\tan \frac{\xi }{2} &=&\sqrt{\frac{1-e_{r}}{1+e_{r}}}\tan \frac{\chi }{2}%
\text{ \ },  \label{xichi} \\
\sin \xi  &=&\frac{\sqrt{1-e_{r}^{2}}\sin \chi }{1+e_{r}\cos \chi }\text{ },
\end{eqnarray}%
with $e_{r}$ in place of the Keplerian eccentricity. Note that in the
Damour-Deruelle formalism \cite{DD}, a different generalized true anomaly
parameter $v$ is introduced$.$ To 1PN accuracy $v$ is related to $\chi $ as%
\begin{equation}
\tan \frac{\chi }{2}=\!\!\left[ 1\!-\!\frac{Gm\mu }{4c^{2}\overline{L}^{2}%
\overline{A}}\left( \!G^{2}m\mu ^{3}\!+\!\frac{12E\overline{L}^{2}}{\mu }%
\!\right) \right] \!\tan \frac{v}{2}\text{ }.  \label{chiv}
\end{equation}%
When using the generalized true anomaly parameter $v$, the equation $\xi
\left( v\right) $ replacing Eq. (\ref{xichi}) will contain the angular
eccentricity$\ e_{\theta }$ rather than $e_{r}$.

\section{Generalized Kepler equation}

The first three terms on the right hand side in the radial equation (\ref%
{rad2}) sum up to:%
\begin{eqnarray}
&&\frac{2E}{\mu }+\frac{2Gm}{r}-\frac{\overline{L}^{2}}{\mu ^{2}r^{2}} 
\notag \\
=\! &&\!\!\left( \frac{\overline{A}}{\overline{L}^{2}}\!+\!\frac{2\delta Q}{%
\mu }\right) \overline{A}\sin ^{2}\chi \!-\!\frac{2\overline{A}}{\mu }\left(
\delta Q+\!\delta P\cos \chi \right) \text{ },  \label{radial}
\end{eqnarray}%
where $\delta P$ and $\delta Q$ are perturbation terms depending on the
physical parameters of the binary and they are given in the Tables \ref%
{deltaP} and \ref{deltaQ}. 
\begin{table}[h]
\caption{Various contributions to $\protect\delta P$.}%
\begin{tabular}{|l|l|}
\hline
$P\!N\!$ & $\!-\frac{Gm\mu ^{2}}{c^{2}\overline{L}^{4}}\left[ \overline{A}%
^{2}(\eta -2)-4G^{2}m^{2}\mu ^{2}\right] $ \\ \hline
$S\!O\!$ & $\!-\frac{G\mu ^{3}}{2c^{2}\overline{L}^{5}}\sum\limits_{i=1,j%
\neq i}^{2}\!\frac{4m_{i}+3m_{j}}{m_{i}}S_{i}\cos \kappa _{i}\left( 
\overline{A}^{2}\!+\!3G^{2}m^{2}\mu ^{2}\right) $ \\ \hline
$S\!S\!$ & $\!\frac{G\mu ^{3}S_{1}S_{2}}{2c^{2}\overline{L}^{6}}\!\left[
\alpha _{SS}\!\left( \overline{A}^{2}\!+\!3G^{2}m^{2}\mu ^{2}\right)
\!+\!\beta _{SS}\!\left( \overline{A}^{2}\!+\!G^{2}m^{2}\mu ^{2}\right) %
\right] $ \\ \hline
$Q\!M\!$ & $\!-\frac{Gm^{3}\mu ^{4}}{4\overline{L}^{6}}\sum_{i=1}^{2}p_{i}%
\biggl[\alpha _{QM}^{i}\left( \overline{A}^{2}+3G^{2}m^{2}\mu ^{2}\right) \!$
\\ 
& $+\beta _{QM}^{i}\left( \overline{A}^{2}+G^{2}m^{2}\mu ^{2}\right) \biggr]$
\\ \hline
$D\!D\!$ & $\!-\frac{\mu ^{3}d_{1}d_{2}}{2\overline{L}^{6}}\left[ \alpha
_{DD}\!\left( \overline{A}^{2}\!+\!3G^{2}m^{2}\mu ^{2}\right) \!+\!\beta
_{DD}\!\left( \overline{A}^{2}\!+\!G^{2}m^{2}\mu ^{2}\right) \right] $ \\ 
\hline
\end{tabular}%
\label{deltaP}
\end{table}
\begin{table}[h]
\caption{Various contributions to $\protect\delta Q$.}%
\begin{tabular}{|l|l|}
\hline
$PN$ & $-\frac{\mu }{8c^{2}\overline{L}^{4}\overline{A}}\bigl [%
2G^{2}m^{2}\mu ^{2}\overline{A}^{2}(3\eta -19)$ \\ 
& $-G^{4}m^{4}\mu ^{4}(\eta +9)+\overline{A}^{4}(3\eta -1)\bigr ]$ \\ \hline
$SO$ & $-\frac{G^{2}m\mu ^{4}}{2c^{2}\overline{L}^{5}\overline{A}}%
\sum\limits_{i=1,j\neq i}^{2}\!\frac{4m_{i}+3m_{j}}{m_{i}}S_{i}\cos \kappa
_{i}\left( 3\overline{A}^{2}+G^{2}m^{2}\mu ^{2}\right) $ \\ \hline
$SS$ & $\frac{G^{2}m\mu ^{4}S_{1}S_{2}}{2c^{2}\overline{L}^{6}\overline{A}}%
\left[ \alpha _{SS}\left( 3\overline{A}^{2}+G^{2}m^{2}\mu ^{2}\right)
+2\beta _{QM}\overline{A}^{2}\right] $ \\ \hline
$QM$ & $-\frac{G^{2}m^{4}\mu ^{5}}{4\overline{L}^{6}\overline{A}}%
\sum_{i=1}^{2}p_{i}\left[ \alpha _{QM}^{i}\left( 3\overline{A}%
^{2}+G^{2}m^{2}\mu ^{2}\right) +2\beta _{SS}^{i}\overline{A}^{2}\right] $ \\ 
\hline
$DD$ & $-\frac{Gm\mu ^{4}d_{1}d_{2}}{2\overline{L}^{6}\overline{A}}\left[
\alpha _{DD}\left( 3\overline{A}^{2}+G^{2}m^{2}\mu ^{2}\right) +2\beta _{DD}%
\overline{A}^{2}\right] $ \\ \hline
\end{tabular}%
\label{deltaQ}
\end{table}
The parametrization (\ref{rchi}) has the advantageous property\thinspace 
\begin{equation}
\frac{dr}{d\chi }=\frac{1}{2}\left( \frac{1}{r_{\min }}-\frac{1}{r_{\max }}%
\right) r^{2}\sin \chi \text{ }.  \label{drdchi}
\end{equation}%
Employing Eqs. (\ref{rchi}) and (\ref{drdchi}) into the radial equation (\ref%
{rad2}), then taking the square root and forming the reciprocal, after a
series expansion to first order in the perturbations we find%
\begin{eqnarray}
\frac{dt}{d\chi }\! &=&\!\frac{\mu r^{2}}{\overline{L}}\!\!\Biggl\{%
\!\!1+\!\!\!\frac{\overline{L}^{2}}{2\mu ^{2}r^{2}\overline{A}^{2}\sin
^{2}\chi }\!\!\biggl [r^{2}\mu A(\delta Q\!+\!\delta P\cos \chi )  \notag \\
&&+2\left( \overline{L}\delta L\!+\!r^{2}\mu \delta E\right)
+\sum_{i=0}^{3}\delta A_{i}r^{2-i}\biggr ]\Biggr\}\text{ },  \label{sing}
\end{eqnarray}%
The integration of this equation is the main purpose of our paper.

Apparently Eq. (\ref{sing}) becomes singular at $\chi =k\pi $, $k\in Z$,
because of the $\sin ^{2}\chi $ term in the denominator. Such singularities
could be just apparent, as was shown for the SO-contribution in \cite{param}%
, We have verified that after forming the common denominator in the bracket,
and inserting the detailed expressions of $\delta P$, $\delta Q$, $\delta L$
and $\delta E$, given in Tables \ref{delL}-\ref{deltaQ}, the numerator
becomes proportional to $\sin ^{2}\chi $. Therefore the singularities are
just apparent rather than real for all type of contributions considered
here. Thus we obtain 
\begin{widetext}
\begin{eqnarray}
\frac{dt}{d\chi }\! &=&\!\frac{\mu r^{2}}{\overline{L}}+\left( \frac{dt}{%
d\chi }\right) _{PN}+\left( \frac{dt}{d\chi }\right) _{SO}+\left( \frac{dt}{%
d\chi }\right) _{SS}+\left( \frac{dt}{d\chi }\right) _{DD}+\left( \frac{dt}{%
d\chi }\right) _{QM}\text{ },  \notag \\
\left( \frac{dt}{d\chi }\right) _{PN}\! &=&\!\frac{\mu r^{2}}{2c^{2}%
\overline{L}^{3}}\Biggl\{\left( \eta -13\right) G^{2}m^{2}\mu ^{2}+\left(
3\eta -1\right) \overline{A}^{2}+(3\eta -8)Gm\mu \overline{A}\cos \chi %
\Biggr\}\text{ },  \notag \\
\left( \frac{dt}{d\chi }\right) _{SO}\! &=&\!-\frac{G\mu ^{2}r^{2}}{2c^{2}%
\overline{L}^{3}}\left( 3Gm\mu +\overline{A}\cos \chi \right)
\sum\limits_{i=1,j\neq i}^{2}\!\frac{4m_{i}+3m_{j}}{m_{i}}S_{i}\cos \kappa
_{i}\text{ },  \notag \\
\left( \frac{dt}{d\chi }\right) _{SS}\! &=&\!\frac{G\mu ^{3}S_{1}S_{2}r^{2}}{%
2c^{2}\overline{L}^{5}}\Biggl\{\left[ Gm\mu \left( 3\alpha _{SS}+2\beta
_{SS}\right) +\overline{A}\left( \alpha _{SS}+\beta _{SS}\right) \cos \chi %
\right] -\frac{2\overline{L}^{4}}{\mu ^{2}r^{2}\overline{A}}\sin \kappa
_{1}\sin \kappa _{2}\cos \!\left[ \chi +2\left( \psi _{0}-\overline{\psi }%
\right) \right] \Biggr\}\text{ },  \notag \\
\left( \frac{dt}{d\chi }\right) _{QM}\! &=&\!-\frac{Gm^{3}\mu ^{4}r^{2}}{4%
\overline{L}^{5}}\sum_{i=1}^{2}p_{i}\Biggl\{\!\left[ Gm\mu \!\left( 3\alpha
_{QM}^{i}\!+\!2\beta _{QM}^{i}\right) +\overline{A}\!\left( \alpha
_{QM}^{i}\!+\!\beta _{QM}^{i}\right) \cos \chi \right] -\frac{2\overline{L}%
^{4}}{\mu ^{2}r^{2}\overline{A}}\sin ^{2}\!\kappa _{i}\cos \!\left[ \chi
+2\left( \psi _{0}\!-\!\psi _{i}\right) \right] \!\Biggr\}\text{ },  \notag
\\
\left( \frac{dt}{d\chi }\right) _{DD}\! &=&\!-\frac{\mu ^{3}d_{1}d_{2}r^{2}}{%
2\overline{L}^{5}}\Biggl\{\left[ Gm\mu \left( 3\alpha _{DD}+2\beta
_{DD}\right) +\overline{A}\left( \alpha _{DD}+\beta _{DD}\right) \cos \chi %
\right] +\frac{2\overline{L}^{4}}{\mu ^{2}r^{2}\overline{A}}\cos \left[ \chi
+2\left( \psi _{0}-\overline{\psi }\right) \right] \Biggr\}\text{ }.
\label{dtdchi}
\end{eqnarray}

\end{widetext}These expressions are regular for any $\chi $.

By integrating Eq. (\ref{dtdchi}) and employing the relation (\ref{xichi})
between the two parametrizations we obtain the generalized Kepler equation (%
\ref{Kepler}). This is the main result of the paper.

\section{The orbital parameters}

In this Section we enlist the detailed expressions of the orbital parameters
appearing either in the eccentric anomaly parametrization (\ref{rxi}) or in
the generalized Kepler equation (\ref{Kepler}).

The semimajor axis is 
\begin{eqnarray}
a_{r} &=&\frac{Gm\mu }{-2E}%
+a_{r}^{PN}+a_{r}^{SO}+a_{r}^{SS}+a_{r}^{QM}+a_{r}^{DD}\text{ },  \notag \\
a_{r}^{PN} &=&\frac{Gm}{4c^{2}}\left( \eta -7\right) \text{ },  \notag \\
a_{r}^{SO} &=&\frac{G\mu }{2c^{2}\overline{L}}\sum\limits_{i=1,j\neq i}^{2}\!%
\frac{4m_{i}+3m_{j}}{m_{i}}S_{i}\cos \kappa _{i}\!\text{ },  \notag \\
a_{r}^{SS} &=&-\frac{G\mu S_{1}S_{2}}{2c^{2}\overline{L}^{2}}\left( \alpha
_{SS}+\beta _{SS}\right) \text{ },  \notag \\
a_{r}^{QM} &=&\frac{Gm^{3}\mu ^{2}}{4\overline{L}^{2}}\sum_{i=1}^{2}p_{i}%
\left( \alpha _{QM}^{i}+\beta _{QM}^{i}\right) \text{ },  \notag \\
a_{r}^{DD} &=&\frac{\mu d_{1}d_{2}}{2\overline{L}^{2}}\left( \alpha
_{DD}+\beta _{DD}\right) \text{ }.  \label{orbparam}
\end{eqnarray}

The radial eccentricity is%
\begin{eqnarray}
e_{r} &=&\frac{\overline{A}}{Gm\mu }%
+e_{r}^{PN}+e_{r}^{SO}+e_{r}^{SS}+e_{r}^{QM}+e_{r}^{DD}\text{ },  \notag \\
e_{r}^{PN} &=&\frac{E}{4c^{2}Gm\mu ^{2}\overline{A}}\left[ \left( 5\eta
-15\right) \overline{A}^{2}-\left( \eta +9\right) G^{2}m^{2}\mu ^{2}\right] 
\text{ },  \notag \\
e_{r}^{SO} &=&\frac{E\left( G^{2}m^{2}\mu ^{2}\!+\!\overline{A}^{2}\!\right) 
}{c^{2}Gm^{2}\mu \overline{L}\overline{A}}\!\sum\limits_{i=1,j\neq i}^{2}\!%
\frac{4m_{i}+3m_{j}}{m_{i}}S_{i}\cos \kappa _{i}\!\text{ },  \notag \\
e_{r}^{SS} &=&-\frac{ES_{1}S_{2}}{c^{2}Gm^{2}\mu \overline{L}^{2}\overline{A}%
}\left[ \left( G^{2}m^{2}\mu ^{2}+\overline{A}^{2}\right) \alpha _{SS}+%
\overline{A}^{2}\beta _{SS}\right] \text{ },  \notag \\
e_{r}^{QM} &=&\frac{Em}{2G\overline{L}^{2}\overline{A}}\sum_{i=1}^{2}p_{i}%
\left[ \left( G^{2}m^{2}\mu ^{2}+\overline{A}^{2}\right) \alpha _{QM}^{i}+%
\overline{A}^{2}\beta _{QM}^{i}\right] \text{ },  \notag \\
e_{r}^{DD} &=&\frac{Ed_{1}d_{2}}{Gm\mu \overline{L}^{2}\overline{A}}\left[
\left( G^{2}m^{2}\mu ^{2}+\overline{A}^{2}\right) \alpha _{DD}+\overline{A}%
^{2}\beta _{DD}\right] \text{ }.
\end{eqnarray}

The mean motion is 
\begin{equation}
n=\frac{2\pi }{T}=\frac{1}{Gm}\left( \frac{-2E}{\mu }\right) ^{3/2}\left[
1-\left( \eta -15\right) \frac{E}{4c^{2}\mu }\right] \text{\ }.
\end{equation}

The time eccentricity is 
\begin{eqnarray}
e_{t} &=&\frac{\overline{A}}{Gm\mu }%
+e_{t}^{PN}+e_{t}^{SO}+e_{t}^{SS}+e_{t}^{QM}+e_{t}^{DD}\text{ },  \notag \\
e_{t}^{PN} &=&-\frac{E}{4c^{2}Gm\mu ^{2}\overline{A}}\left[ \left( 7\eta
-17\right) \overline{A}^{2}+\left( \eta +9\right) G^{2}m^{2}\mu ^{2}\right] 
\text{ },  \notag \\
e_{t}^{SO} &=&\frac{EG\mu }{c^{2}\overline{L}\overline{A}}%
\!\sum\limits_{i=1,j\neq i}^{2}\!\frac{4m_{i}+3m_{j}}{m_{i}}S_{i}\cos \kappa
_{i}\text{ },  \notag \\
e_{t}^{SS} &=&-\frac{ES_{1}S_{2}G\mu \alpha _{SS}}{c^{2}\overline{L}^{2}%
\overline{A}}\text{ },  \notag \\
e_{t}^{QM} &=&\frac{EGm^{3}\mu ^{2}}{2\overline{L}^{2}\overline{A}}%
\sum_{i=1}^{2}p_{i}\alpha _{QM}^{i}\text{ },  \notag \\
e_{t}^{DD} &=&\frac{E\mu d_{1}d_{2}}{\overline{L}^{2}\overline{A}}\alpha
_{DD}\text{ }.
\end{eqnarray}

In what follows, we enlist the parameters $f_{t}$ and $f_{t}^{i}$, the
analogues of which also appear in the extension to 2PN \cite{Damour-Schafer}
of the Damour-Deruelle parametrization, however have no PN counterpart. The
parameter $f_{t}$ receives only SS and DD type contributions%
\begin{eqnarray}
f_{t} &=&f_{t}^{SS}+f_{t}^{DD}\text{ },  \notag \\
f_{t}^{SS} &=&-\left( \frac{-2E}{\mu }\right) ^{3/2}\frac{\mu S_{1}S_{2}}{%
c^{2}m\overline{A}\overline{L}}\sin \kappa _{1}\sin \kappa _{2}\text{ }, 
\notag \\
f_{t}^{DD} &=&\left( \frac{-2E}{\mu }\right) ^{3/2}\frac{\mu d_{1}d_{2}}{Gm%
\overline{A}\overline{L}}\text{ }.
\end{eqnarray}%
Finally the parameters $f_{t}^{i}$, originating from the QM interaction, are
given as%
\begin{equation}
f_{t}^{i}=\left( \frac{-2E}{\mu }\right) ^{3/2}\frac{m^{2}\mu ^{2}}{2%
\overline{A}\overline{L}}p_{i}\sin ^{2}\kappa _{i}\text{ \ }.
\label{orbparamend}
\end{equation}

\section{Concluding remarks}

The generalized Kepler equation (\ref{Kepler}) with the orbital parameters (%
\ref{orbparam})-(\ref{orbparamend}), together with any of the
parametrizations (\ref{rxi}) or (\ref{rchi}) and their relation (\ref{xichi}%
) represent the \textit{complete solution of the radial motion of the
compact binary on eccentric orbit, to linear order in the perturbations}.
All perturbations arising from relativistic corrections and from the
presence of spins, mass quadrupole and magnetic dipole moments are included
here to linear order (PN, SO, SS, QM, DD contributions).

The generalized Kepler equation contains two parameters, the generalized
eccentric anomaly $\xi $ (this is defined similarly as the parameter $u$ in 
\cite{DD,Damour-Schafer} and the generalized true anomaly $\chi $ (different
from the parameter $v$ of \cite{DD,Damour-Schafer}, their relation being
given to 1PN accuracy by Eq. (\ref{chiv})).

The generalization of the relation (\ref{chiv}) to 2PN accuracy would imply
to give the $\chi -$parametrization (\ref{rchi}) to 2PN, however by the
method described in \cite{param} $r\left( \chi \right) $ can be defined only
to linear order in the perturbations. Nevertheless, the linear contributions
to the Kepler equation included in the present paper (containing the
parameters $\xi $ and $\chi )$, and the 2PN\ contributions \cite%
{Damour-Schafer} (containing $u\equiv \xi $ and $v$) can be simply summed
up, at the price of having all three parameters ($\xi \equiv u$, $\chi $ and 
$v$) present in the formalism. Notably at 2PN orders a new parameter, $g_{t}$
is also present \cite{WS}. We note here that the 3PN contribution to the
Kepler equation is also known \cite{KG}, however the contribution of the
first PN correction of the SO interaction, arising at 2.5 PN is not.
Therefore we conclude that for the moment, the radial motion is solved only
to 2PN orders accuracy.

In contrast with the PN and 2PN Kepler equations, our Eq. (\ref{Kepler})
contains the additional angles $\psi _{0}$ and $\psi _{i}$. During one
radial period these can be considered constants \cite{GPV}. On the long run
however, all these angles slowly vary as the orbit and the spins undergo
precessional motions. In order to describe the slow evolution of these
angles, the study of the angular part of the motion (as opposed to the
radial one) is necessary, with the inclusion of all perturbations, to linear
order. This is available for the PN perturbation \cite{DD} and SO
perturbation \cite{KG}, however the latter holds only for special cases
(equal masses or a single spin). A systematic investigation of the angular
part of the perturbed Keplerian motion is under way \cite{KMG2}.

We remark that the circular orbit limit of our formulae should \textit{not}
be taken as $\overline{A}\rightarrow 0$. This is because the standard
interpretation of the Laplace-Runge-Lenz vector holds only in the Newtonian
limit. In the circular orbit limit all corrections considered here add
nonvanishing contributions to $\overline{A}$.

Our generic Kepler equation (\ref{Kepler}) and the orbital elements (\ref%
{orbparam})-(\ref{orbparamend}) correctly reproduce the PN contributions 
\cite{DD} and the SO contributions \cite{KG}. However all of the other
contributions are new.

\section{Acknowledgements}

This work was supported by OTKA grants no. T046939 and TS044665. L.\'{A}.G.
was further supported by the J\'{a}nos Bolyai Scholarship of the Hungarian
Academy of Sciences.


\begin{thebibliography}{99}
\bibitem{LIGO} A. Abramovici et al., Science 256, 325 (1992).

\bibitem{VIRGO} C. Bradaschia et al., Nucl. Instrum. Methods A 289, 518
(1990).

\bibitem{GEO} J. Hough, in Proceedings of the Sixth Marcell Grossmann
Meeting, edited by H. Sato and T. Nakamura (World Scientific, Singapore,
1992), p. 192.

\bibitem{Tama} K. Kuroda et al., in Proceedings of International Conference
on Gravitational Waves: Sources and Detectors, eds. I. Ciufolini and F.
Fidecaro (World Scientific, Singapore, 1997), p. 100.

\bibitem{LIGO1} LIGO Scientific Collaboration: B. Abbott, et al., \textit{%
Search for gravitational waves from galactic and extra--galactic binary
neutron stars }gr-qc/0505042 (2005).

\bibitem{LIGO2} LIGO Scientific Collaboration: B. Abbott, et al., \textit{%
Search for Gravitational Waves from Binary Black Hole Inspirals in LIGO Data,%
} gr-qc/0509129 (2005).

\bibitem{IBBH} P.R. Brady, J.D.E. Creighton, and K.S. Thorne \prd\textbf{58}%
, 061501 (1998).

\bibitem{Buonanno1} A. Buonanno, Y. Chen, and M. Vallisneri, \prd\textbf{67}%
, 024016 (2003).

\bibitem{3PN} L. Blanchet, B. R. Iyer, B. Joguet, \prd\textbf{65} 064005
(2002); Erratum-ibid. \prd\textbf{71},129903 (2005).

\bibitem{BOC} B. M. Barker and R. F. O'Connell, \prd\textbf{2}, 1428 (1970).

\bibitem{OC} R. F. O'Connell, Phys. Rev. Lett. \textbf{93,} 081103 (2004).

\bibitem{Burgay} N. Burgay et al., Nature \textbf{426}, 531 (2003).

\bibitem{Lyne} A. G. Lyne et al., Science \textbf{303}, 1153 (2004).

\bibitem{ACST} A. Apostolatos, C. Cutler, G. J. Sussman, and K. S. Thorne, %
\prd\textbf{49}, 6274 (1994).

\bibitem{KG} C. K\"{o}enigsd\"{o}rffer, A. Gopakumar, \prd\textbf{71,}
024039 (2005). For the conversion to our notations we remark that $2\mathbf{S%
}_{eff}\cdot \mathbf{L}=4\mathbf{L\cdot S}+3\mathbf{L\cdot \sigma }$.

\bibitem{KG2} C. K\"{o}enigsd\"{o}rffer, A. Gopakumar, \textit{Parametric
derivation of the observable relativistic periastron advance for binary
pulsars, }gr-qc/0509012 (2005).

\bibitem{GIKB} P. Grandclement, M. Ihm, V. Kalogera, K. Belczynski, \prd%
\textbf{69,} 102002, (2004).

\bibitem{RS} R. Rieth and G. Schafer, Class. Quantum Grav \textbf{14}, 2357,
(1997).

\bibitem{GPV} L. \'{A}. Gergely, Z. I. Perj\'{e}s, and M. Vas\'{u}th, Phys.
Rev D \textbf{57}, 876 (1998);

L. \'{A}. Gergely, Z. I. Perj\'{e}s, and M. Vas\'{u}th, Phys. Rev D \textbf{%
57}, 3423 (1998);

L. \'{A}. Gergely, Z. I. Perj\'{e}s, and M. Vas\'{u}th, Phys. Rev D \textbf{%
58}, 124001 (1998).

\bibitem{BD99} A. Buonanno, T. Damour, \prd\textbf{59} 084006 (1999).

\bibitem{BD00} A. Buonanno, T. Damour, \prd\textbf{62} 064015(2000).

\bibitem{BCD05} A. Buonanno, Y. Chen, T. Damour, \textit{Transition from
inspiral to plunge in precessing binaries of spinning black holes, }%
gr-qc/0508067 (2005).

\bibitem{BCV2} A. Buonanno, Y. Chen, and M. Vallisneri, \prd\textbf{67,}
104025 (2003).

\bibitem{BCV2P} A. Buonanno, Y. Chen, Y. Pan, H. Tagoshi, M. Vallisneri, 
\textit{Detecting gravitational waves from precessing binaries of spinning
compact objects. II. Search implementation for low-mass binaries, }%
gr-qc/0508064 (2005).

\bibitem{spinspin} L. \'{A}. Gergely, Phys. Rev. D \textbf{61}, 024035
(2000);

L. \'{A}. Gergely, Phys. Rev D \textbf{62}, 024007 (2000).

\bibitem{quadrup} L. \'{A}. Gergely, Z. Keresztes, \prd {\bf 67}, 024020
(2003).

\bibitem{mdipol} M. Vas\'{u}th, Z. Keresztes, A. Mih\'{a}ly, and L. \'{A}.
Gergely, \prd {\bf 68}, 124006 (2003).

\bibitem{MVG} B. Mik\'{o}czi, M.Vas\'{u}th, L. \'{A}. Gergely, \prd\textbf{%
71,} 124043 (2005).

\bibitem{SSC1} M. H. L. Pryce, Proc. R. Soc. A \textbf{195}, 62 (1948).

\bibitem{SSC2} T. D. Newton, E. P. Wigner, Rev. Mod. Phys. \textbf{21}, 400
(1949).

\bibitem{DD} T. Damour and N. Deruelle, {\ Ann. Inst. Henri Poincar\'{e}}\ A 
\textbf{43 }, 107 (1985).

\bibitem{KWW} L. Kidder, C. Will, and A. Wiseman, Phys.Rev. D \textbf{47},
4183 (1993).

\bibitem{Poisson} E. Poisson, \prd {\bf 57}, 5287 (1998).

\bibitem{IT} K. Ioka and T. Taniguchi, \apj  {\bf 537}, 327 (2000).

\bibitem{Kidder} L. Kidder, \prd   {\bf 52}, 821 (1995).

\bibitem{param} L. \'{A}. Gergely, Z. I. Perj\'{e}s, and M. Vas\'{u}th,
Astrophys. J. Suppl. \textbf{126}, 79 (2000).

\bibitem{Damour-Schafer} T. Damour, G. Sch\"{a}fer, Nouvo Cimento B \textbf{%
101}, 127 (1988).

\bibitem{WS} G. Sch\"{a}fer, N. Wex, Phys. Lett A \textbf{174}, 196 (1993),
erratum: \textbf{177}, 461 (1993).

\bibitem{KMG2} Z. Keresztes, B. Mik\'{o}czi, and L. \'{A}. Gergely, in
preparation.
\end{thebibliography}
\end{document}